\documentclass[aps,prl,twocolumn,epsfig,pdflatex]{revtex4-1}

\usepackage{amsmath,amssymb}
\usepackage{graphicx}

\usepackage{float}
\usepackage[usenames,dvipsnames]{xcolor}
\usepackage{amsthm,comment}
\usepackage{booktabs}
\usepackage[export]{adjustbox}
\usepackage[section]{placeins}
\usepackage[caption=false]{subfig}
\usepackage[percent]{overpic}
\bibpunct{[}{]}{,}{n}{}{}
\definecolor{red1}{HTML}{FF4136}
\definecolor{green1}{HTML}{00802b}

\usepackage[hidelinks]{hyperref} 

\hypersetup{colorlinks=true,citecolor=Red,linkcolor=Blue,urlcolor=Blue}
\usepackage[export]{adjustbox}

\graphicspath{{Figures/}}

\begin{document}

\title {Excitonic wave-packet evolution in a two-orbital Hubbard model chain: \\
A real-time real-space study} 
\author{Bradraj Pandey$^{1,2}$,  
Gonzalo Alvarez$^{3}$, and Elbio Dagotto$^{1,2}$}
\affiliation{$^1$Department of Physics and Astronomy, University of Tennessee, Knoxville, Tennessee 37996, USA \\ 
$^2$Materials Science and Technology Division, Oak Ridge National Laboratory, Oak Ridge, Tennessee 37831, USA \\ 
$^3$Computational Sciences $\&$ Engineering Division and Center for Nanophase Materials Sciences,
Oak Ridge National Laboratory,~Oak Ridge,~Tennessee 37831,~USA
 }

\begin{abstract}

Motivated by experimental developments introducing the concept of spin-orbit separation, 
	we study the real-space time evolution of an excitonic 
	wave-packet using a two-orbital Hubbard model. The exciton 
 is created by exciting an electron from a lower energy half-filled 
	orbital to a higher energy empty orbital. We carry out the real-time dynamics
	of the resulting excitonic wave-packet, using the time-dependent density matrix 
	renormalization group method. 
	We find clear evidence of charge-spin and spin-orbit separation 
	in real-space, by tracking the time evolution of 
	local observables. We show that the velocity of the orbiton can be 
	tuned by varying the inter-orbital interactions. 
	 We also present a comparative study
	of a hole (in one orbital)  and exciton (in two orbitals) 
	dynamics in one-dimensional systems.
	Moreover, we analyze the dynamics of an exciton with spin-flip excitation, 
	where we observe fractionalized spinons induced by Hund’s interaction.

\end{abstract}


\maketitle

{\it Introduction.} The dynamics of excitations in low-dimensional 
compounds has attracted considerable attention~\cite{kim,tokura}.
Experimentally observed excitations include holons, spinons, 
doublons, and excitons~\cite{koh,matiks}. In particular, the study of 
excitons in multiband insulators unveiled interesting surprises~\cite{zhang,rincon}. 
Localized and delocalized charge-transfer excitons 
were observed experimentally in La$_2$CuO$_4$ and La$_2$NiO$_4$, 
respectively~\cite{shukla}. 
More recently, spin-orbit excitons were observed in Sr$_2$IrO$_4$ 
using Resonant Inelastic  X-ray Scattering (RIXS)~\cite{casa}. 
The excitonic dynamic in Sr$_2$IrO$_4$ was described as analogous 
to the propagation of holes in a cuprate's antiferromagnetic (AFM) background~\cite{van}.

Due to reduced dimensionality and strong correlation effects,
quasi one-dimensional (1D) systems display 
exotic dynamical properties~\cite{giamarchi,halden,pandey}, including the fractionalization of 
low-energy excitations~\cite{kim,koh} into spin (spinons) 
and charge (holons) excitations propagating with different velocities~\cite{voit,jagla,julian}.
The existence of spin-charge separation was shown experimentally
early on in quasi one-dimensional systems~\cite{yacoby,kollath}. 
Interestingly, the fractionalization of electronic excitations
 is not limited only to spin and charge, but it can also
include the orbital degree of freedom.
In fact, spin-orbital separation was observed 
experimentally in the transition metal compound Sr$_2$CuO$_3$~\cite{zhou}.
Using high-resolution RIXS experiments, spin-orbital separation was
also observed in the ladder system CaCu$_2$O$_3$~\cite{monney}. 

Recently, the spin-orbit separation in Mott insulating systems was studied theoretically 
using the effective Kugel-Khomskii model~\cite{daghofer}. In the limit of vanishing Hund's coupling,
the propagation of the orbiton in a ferro-orbital and antiferro-magnetic chain was shown to
map into a ``single hole'' moving in an AFM chain with its dynamics described by an 
effective $t-J$ model~\cite{daghofer}.
However, in transition metal compounds, the Hund interaction
plays an important role and, depending on the material, it can be strong.
Precisely for a strong Hund's coupling, 
the propagation of an orbiton, with inter-orbital FM  and AFM spin alignments 
(in the excited states of the superexchange process) are not equal, 
  and the simple mapping between orbiton and $t-J$ hole dynamics is
no longer valid~\cite{hever,betto}. 
More recently, using RIXS the impact of the Hund's
interaction on the orbiton propagation of the quasi-1D AFM compound
Ca$_2$CuO$_2$ was studied~\cite{betto}. It was observed
that robust Hund's interactions are required in the theoretical
description to understand the experimental orbital spectrum~\cite{betto}.

In this letter, we provide the first study of spin-orbit
separation in a real-time and real-space formalism by creating a finite momentum 
excitonic wave-packet at time $t=0$,
using a one-dimensional chain with two orbitals at each site. 
This wave-packet is created by exciting an electron
from a half-filled orbital to an empty higher-energy orbital, as in experiments.
Previous studies primarily focused on the spectral properties to study spin-orbit separation and for simplicity relied on Kugel-Khomskii models in
the strong coupling limit~\cite{daghofer,hever}. 
Here we consider a more general
multi-orbital Hubbard Hamiltonian at intermediate coupling strengths, 
accounting also for charge fluctuations and with focus on the influence of the 
Hund's coupling. 
To study the excitonic real-time dynamics, we use the time-dependent
density-matrix-renormalization group (t-DMRG) method~\cite{white,feiguin}.
We have observed that after creating the exciton, 
the hole (in the half-filled orbital) and the electron
(in the empty orbital)
always move together, while the spin wave-packet
in the half-filled orbital independently evolves from the charge wave-packet.
We also compare the dynamics of
a hole in a one-orbital chain versus the dynamics of an exciton in
a two-orbital chain. Overall, at intermediate coupling and 
for robust values of the Hund's interaction, we find clear evidence of
spin-orbit separation as time grows.
Moreover, we quantitatively study the relation 
between the Hund's coupling and the orbiton velocity, finding
that this orbiton's velocity increases with an increase in
the Hund's coupling magnitude, while the spinon's velocity remains
unaffected. Furthermore, we also present the dynamics of a spin-flip exciton (the previous 
discussion was for a spin preserving exciton),
where we find fractionalized spinons, induced by the strong Hund's coupling.

{ \it Model and Method.} 
We use the two-orbital Hubbard model on a chain. The model can be written 
as the sum of kinetic and interaction energy terms $H=H_k+H_{in}$~\cite{luo}. 
The kinetic (tight-binding) portion contains the nearest-neighbor hopping along the 
chain direction defined as:
\begin{equation}
	H_k = -t_{hop}\sum_{\langle i j \rangle,\sigma,\gamma} \left(c^{\dagger}_{i\sigma\gamma}c^{\phantom\dagger}_{j \sigma \gamma}+H.c.\right) + \sum_{i,\gamma, \sigma} \Delta_{\gamma} n_{i \sigma \gamma}
\end{equation}

\noindent where $c^{\dagger}_{i\sigma\gamma}$ creates an electron at the chain site $i$, with spin $z$-axis projection $\sigma$, and
on orbital $\gamma$ (either orbital $a$ or $b$). $t_{hop}$ is the hopping integral. 
For simplicity, we considered only intra-orbital hoppings  
along the chain and we used identical hopping for both orbitals [$t_a=t_b=t_{hop}=1$]. $\Delta_{\gamma}$ denotes the crystal field term and $n_{i \sigma  \gamma}$ is the orbital-resolved number operator at site $i$.
We  fix the crystal-field parameters as $\Delta_a=4.1$ and $\Delta_b=0$. 
The large crystal field $\Delta_a=4.1 \gtrsim 4 t_{hop}$
ensures only orbital $b$ is occupied in the non-interacting ground state because the bandwidth $W$ of
orbital $a$ is $W=4t_{hop}$.

The electronic interaction portion is canonical:
\begin{eqnarray}
H_{in}= U\sum_{i,\gamma}n_{i\uparrow \gamma} n_{i\downarrow \gamma} +\left(U'-\frac{J_H}{2}\right) \sum_{i,\gamma < \gamma'} n_{i \gamma} n_{i\gamma'} \nonumber \\
-2J_H  \sum_{i,\gamma < \gamma'} {{\bf S}_{i\gamma}}\cdot{{\bf S}_{i\gamma'}}+J_H  \sum_{i,\gamma < \gamma'} \left(P^{\dagger}_{i\gamma} P^{\phantom\dagger}_{i\gamma'}+H.c.\right). 
\end{eqnarray}
The first term is the on-site Hubbard repulsion between 
$\uparrow$ and $\downarrow$ electrons in the same orbital. 
The second term is the electronic repulsion between electrons at 
different orbitals. The standard relation $U'=U-2J_H$ is here assumed. 
The third term is the 
ferromagnetic Hund's interaction between electrons occupying the active 
two orbitals $\gamma={a,b}$ of the same site. ${\bf S}_{i\gamma}$ is 
the total spin of orbital $\gamma$ at site $i$. The last term is the 
pair-hopping between different orbitals,
where $P_{i \gamma}$=$c_{i \downarrow \gamma} c_{i \uparrow \gamma}$.
\begin{figure}[h]
\centering
\rotatebox{0}{\includegraphics*[width=\linewidth]{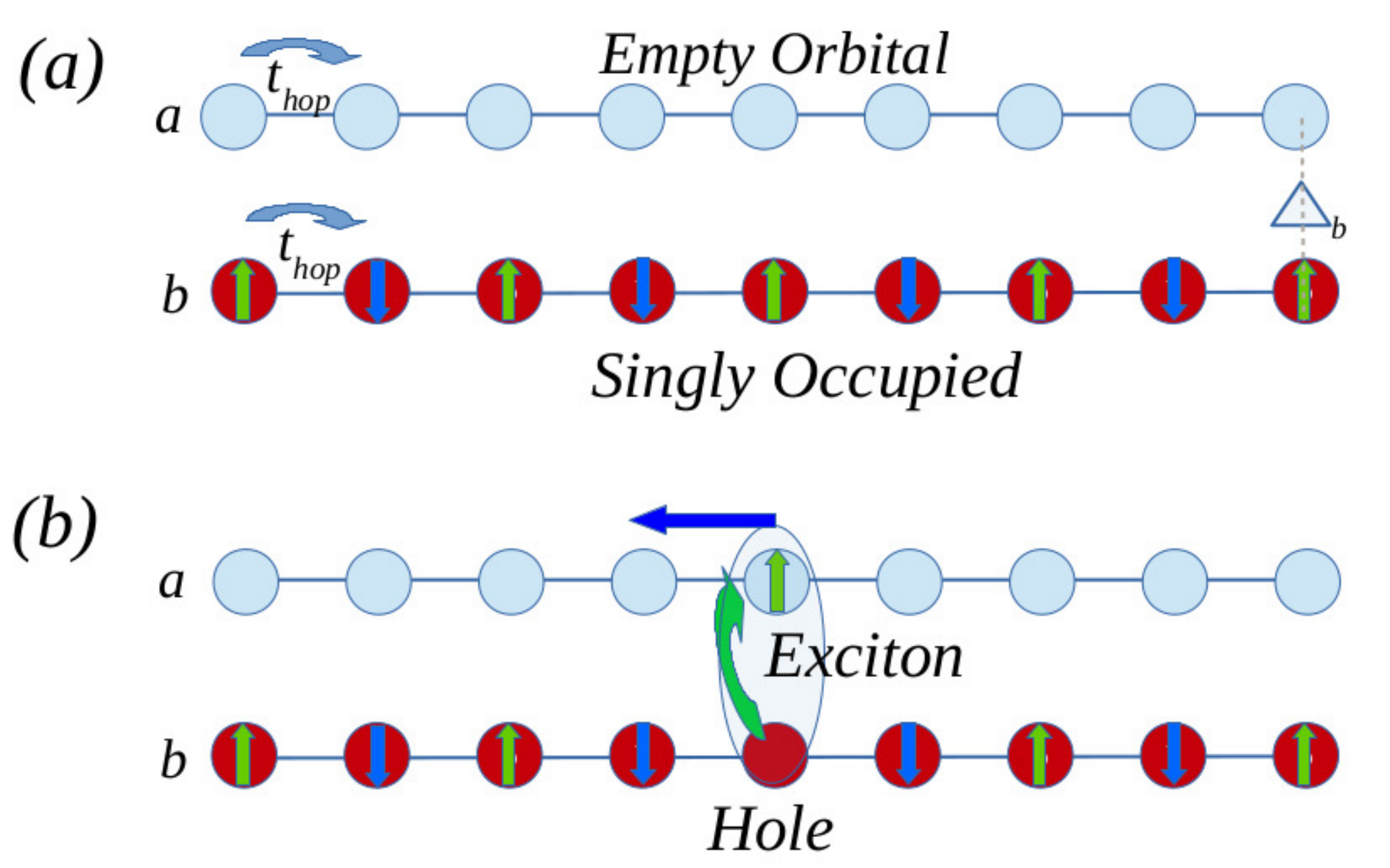}}
	\caption{Schematic representation of a one-dimensional chain with two orbitals ($a$ and $b$) at each site. The blue circle represents orbital $a$ and the red circle orbital $b$. 
	(a) The orbitals are separated by a large crystal field $\Delta$.
	Orbital $b$ is half-filled, whereas orbital $a$ is empty.
	(b) At $t=0$ an exciton is created by exciting an electron from
	the half-filled orbital $b$ (\emph{i.~e.}, one electron per site in a staggered spin pattern) to orbital $a$ which is empty. The exciton has a finite momentum $k_0$ indicated by the blue arrow.
	}
\label{Fig1}
\end{figure}

To obtain the ground state $|\Psi_0\rangle $ of this model,
we employed the static DMRG method. For our numerical calculations,
 we use a system size $L= 36$ with two orbitals at each site and 
 we kept $m=1200$ states. The exciton Gaussian wave-packet is created with 
spin $\sigma$ and a crystal momentum $k_0$, by applying the operator
\begin{equation}
	h^{\dagger}_{\sigma}(k_0) = A \sum_j e^{-\left(j-j_0\right)^2/2\omega_r^2}e^{-ik_0j} c^{\dagger}_{j\sigma a}c^{\phantom\dagger}_{j \sigma b}
\end{equation}
\noindent to the ground state $|\Psi_0\rangle $. This operator excites an
electron from the half-filled orbital $b$ to the empty orbital $a$, 
centered at site $i_0=18$ and with width $\omega_r = 2.54$. The number 2.54 implies 
the size of the initial Gaussian is exactly six lattice spacings at half height, a size that we considered
adequate for easy visualization. 
$A$ is the normalization constant of the Gaussian wave-packet.
Due to the finite width $\omega_k=1/2\pi \omega_r$
of the Gaussian wave-packet in momentum space, we fix the crystal momentum  
 at $k_0=-0.5\pi +4 \omega_k$ (\emph{i.~e.}, close to the highest occupied electronic
 level with width $\omega_k=0.06$). Because we construct a wave packet with 
 a net nonzero momentum [${\bold{k_0}}(exciton) = {\bold{k_e}}(electron) +
 {\bold{k_h}}(hole)$ at time $t=0$] that points in our case towards the left, 
 the resulting time evolution will {\it not} be left-right symmetric. 

We investigate numerically the time evolution of the one-exciton state 
$|\Psi_e \rangle = h^{\dagger}_{\sigma}(k_0) |\Psi_0\rangle $ under the influence of
 $H$ i.e. $|\Psi (t)\rangle= e^{-iHt} |\Psi_e\rangle$. To perform the time evolution,
 we have implemented the Krylov space decomposition in the DMRG
 code~\cite{alvarez,gonzalo}. For the DMRG calculations,
 at least 1200 states were kept during the time evolution.
\begin{figure}[h]
\centering
\rotatebox{0}{\includegraphics*[width=\linewidth]{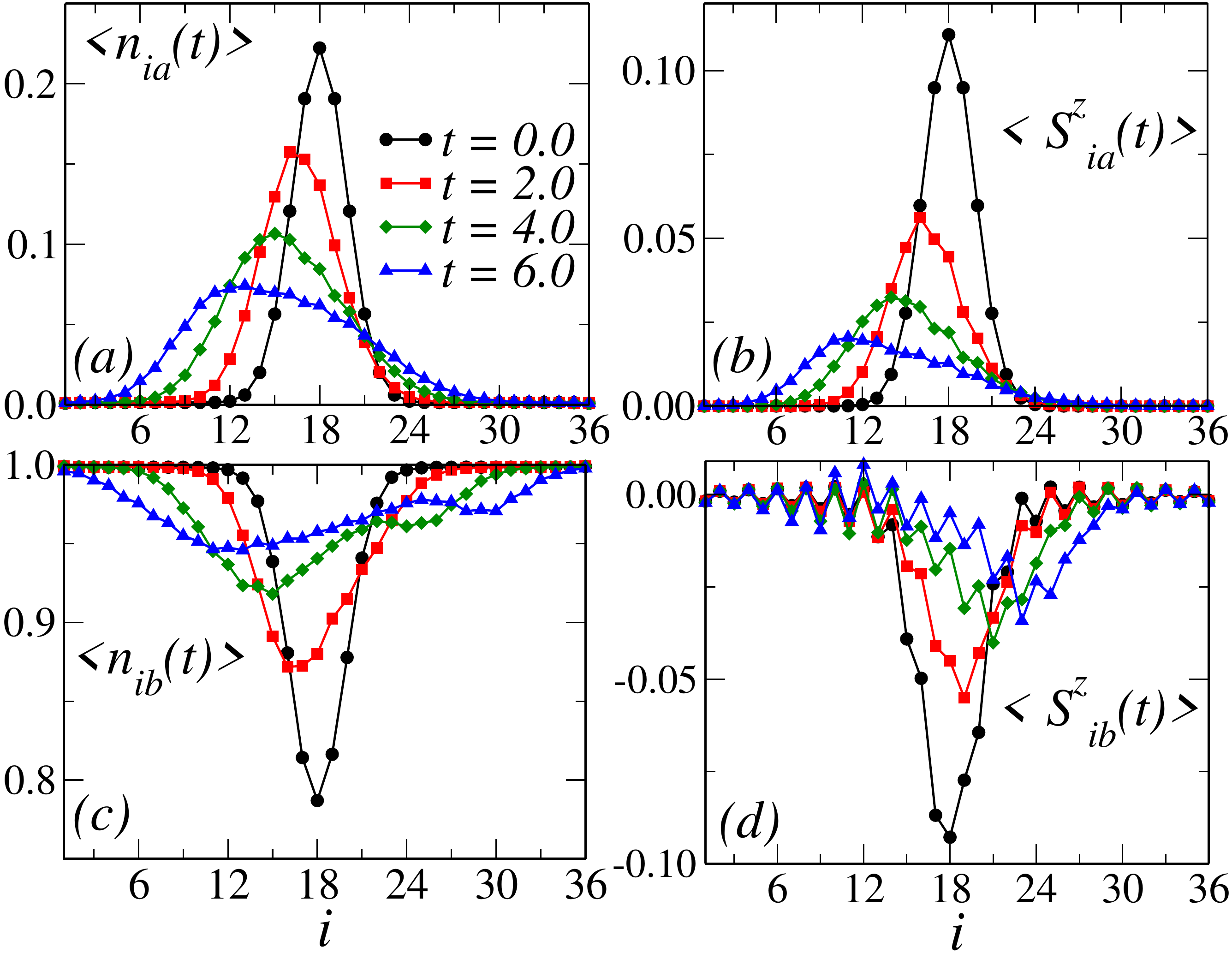}}
\rotatebox{0}{\includegraphics*[width=\linewidth]{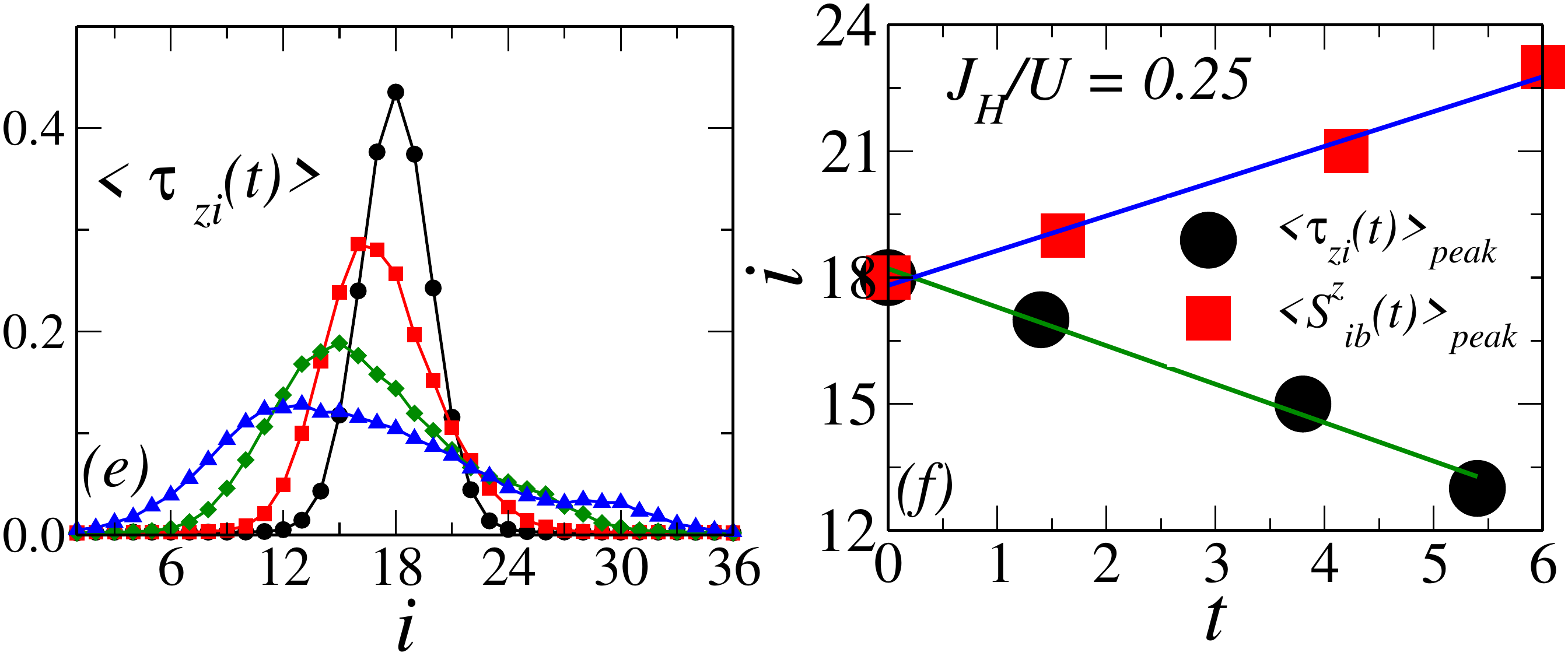}}
	\caption{
	Snapshots of the evolution of wave-packets at different times: 
	(a) charge density $\langle n_{ib}(t) \rangle$, 
	(b) charge density $\langle n_{ia}(t) \rangle$, (c) spin density 
	$\langle S^z_{ib}(t) \rangle$, (d) spin density $\langle S^z_{ia}(t)\rangle$, and      
        (e) orbital density  $\langle \tau_{zi}(t) \rangle$. 
	At $t=0$, wave-packets are at the center of the system i.e. 
	site $i_0=18$. (e) Positions of the peaks of the orbital 
	(black circles) and spin (red squares) wave-packets vs. $t$. 
	 The peak positions are fitted with  straight lines to extract 
	 the velocity of orbital and spin wave-packets.
	These results were obtain at $J_H/U=0.25$ and $U/W=1.0$ using t-DMRG 
	for a $L=36$ sites system.
	}
\label{Fig2}
\end{figure}

 To study the dynamics of the excitonic wave-packet,
 we measure the following observables at each time step:
\begin{eqnarray}
\langle n_{ia}(t) \rangle = \langle \Psi(t)| n_{ia\uparrow}+ n_{ia\downarrow} | \Psi(t) \rangle \\ 
\langle n_{ib}(t) \rangle = \langle \Psi(t)| n_{ib\uparrow}+ n_{ib\downarrow} | \Psi(t) \rangle \\
\langle S^z_{ia}(t) \rangle = \langle \Psi(t)|\left(n_{ia\uparrow}- n_{ia\downarrow}\right)/2 | \Psi(t) \rangle \\
\langle S^z_{ib}(t) \rangle = \langle \Psi(t)|\left(n_{ib\uparrow}- n_{ib\downarrow}\right)/2 | \Psi(t) \rangle \\
\langle \tau_{zi}(t) \rangle = \langle \Psi(t)| n_{ia}- n_{ib} | \Psi(t) \rangle
\end{eqnarray}
\noindent where $\langle n_{ia}(t) \rangle$ and $\langle n_{ib}(t) \rangle$ are the orbital-resolved time dependent charge densities of orbitals $a$ and $b$. 
$\langle S^z_{ia}(t) \rangle$ and $\langle S^z_{ib}(t) \rangle$ are the respective orbital-resolved $z$-component of the time dependent spin densities. 
 $\langle \tau_{zi} (t) \rangle$ is the $z$-component of the time-dependent orbital density.
All these quantities are site dependent. 

{ \it Results.}
The Hamiltonian ground state, at overall quarter-filling ($L=36$ and total number of electrons
$N_e=36$)
and parameters $U/W=1.0$, $J_H/U=0.25$~\cite{hund}, and $\Delta_a=4.1$, 
results in a situation where orbital $b$ is a half-filled Mott-insulator 
with AFM-spin correlations, while orbital $a$ remains empty. 
At time $t=0$, the process previously described
 leads to an exciton centered in the middle of the chain, i.e. at site $i_0=18$.
This results in a  hole wave-packet $\langle n_{ib}(t) \rangle$ 
in orbital $b$ [Fig.~\ref{Fig2}(a)] and an electron wave-packet $\langle n_{ia}(t) \rangle$ 
in orbital $a$ [Fig.~\ref{Fig2}(c)].  The excitation of an electron from orbital $b$ at $t=0$  also
creates spin-excitations $\langle S^z_{ia}(t) \rangle$ with up spins in orbital $a$ [Fig.~\ref{Fig2}(b)] and
down spins in orbital $b$ $\langle S^z_{ib}(t) \rangle$ [Fig.~\ref{Fig2}(d)].

As shown in Figs.~\ref{Fig2}(a) and (b), 
with increasing time the charge wave-packet
 $\langle n_{ia}(t) \rangle$ and the spin wave-packet $\langle S^z_{ia}(t) \rangle$ 
 at orbital $a$ (the originally empty orbital) move with similar speeds toward the left from the 
 central site $i_0=18$, indicating no spin-charge separation 
for orbital $a$, as expected for an electron moving in an empty medium. 
 Interestingly, the charge wave-packets $\langle n_{ia}(t) \rangle$
 and $\langle n_{ib}(t) \rangle$ move together, as mirror images of
 each other [see Figs.~\ref{Fig2}(a) and (c)]. 
 The reason is that the inter-orbital interaction $U'=U-2J_H$ acts as an 
effective attraction between the hole in orbital $b$ and electron 
in orbital $a$~\cite{yang}. Intuitively, when the hole of orbital $b$
and the electron in orbital $a$ are in the same site,
the strong inter-orbital repulsion
energy $U'$ is not active (as compared to the case where two electrons are on the same site).
This results in the formation of an electron-hole bound pair exciton which
moves together with increasing time $t$ towards the left of site $i_0=18$.
The charge wave-packet $\langle n_{ib}(t) \rangle$ 
and spin wave-packet $\langle S^z_{ib}(t) \rangle$ move in opposite 
directions with time, providing clear evidence 
of spin-charge separation [see Fig.~\ref{Fig2}(c) and (d)],
in the half-filled orbital $b$.

In the electron-hole pair exciton, the electron promoted from the
half-filled orbital $b$ to the unoccupied orbital $a$ is also equivalent 
to creating an {\it orbiton}~\cite{lane,hever}. In Fig.~\ref{Fig2}(e), 
we show the orbiton dynamics via $\langle \tau_{zi}(t) \rangle$ evolving with time $t$. 
The orbital wave-packet $\langle \tau_{zi}(t) \rangle$ moves similarly to 
$\langle n_{ib}(t) \rangle$ and $\langle n_{ia}(t) \rangle$, towards the left
 form the central site $i_0=18$, while the spin wave-packet 
 $\langle S^z_{ib}(t) \rangle$ moves toward the right. Thus, our result can be reinterpreted 
as a signature of spin-orbit separation in real-space 
with increasing time $t$. To determine the velocities of the orbital
and spin excitations, we monitored the positions of the peak values of
$\langle \tau_{zi}(t) \rangle$ and $\langle S^z_{ib}(t) \rangle$
vs. time. Using simple linear fits to extract the orbiton ($v_{\tau}$) and
spinon ($v_s$) velocities [Fig.~\ref{Fig2}(f)], we find that the orbital 
wave-packet ($v_{\tau}=-0.91$) has a speed only slight faster than the spin 
wave-packet ($v_s=0.82$), at $J_H/U=0.25$ and $U/W=1.0$. 
\begin{figure}[h]
\centering
\rotatebox{0}{\includegraphics*[width=\linewidth]{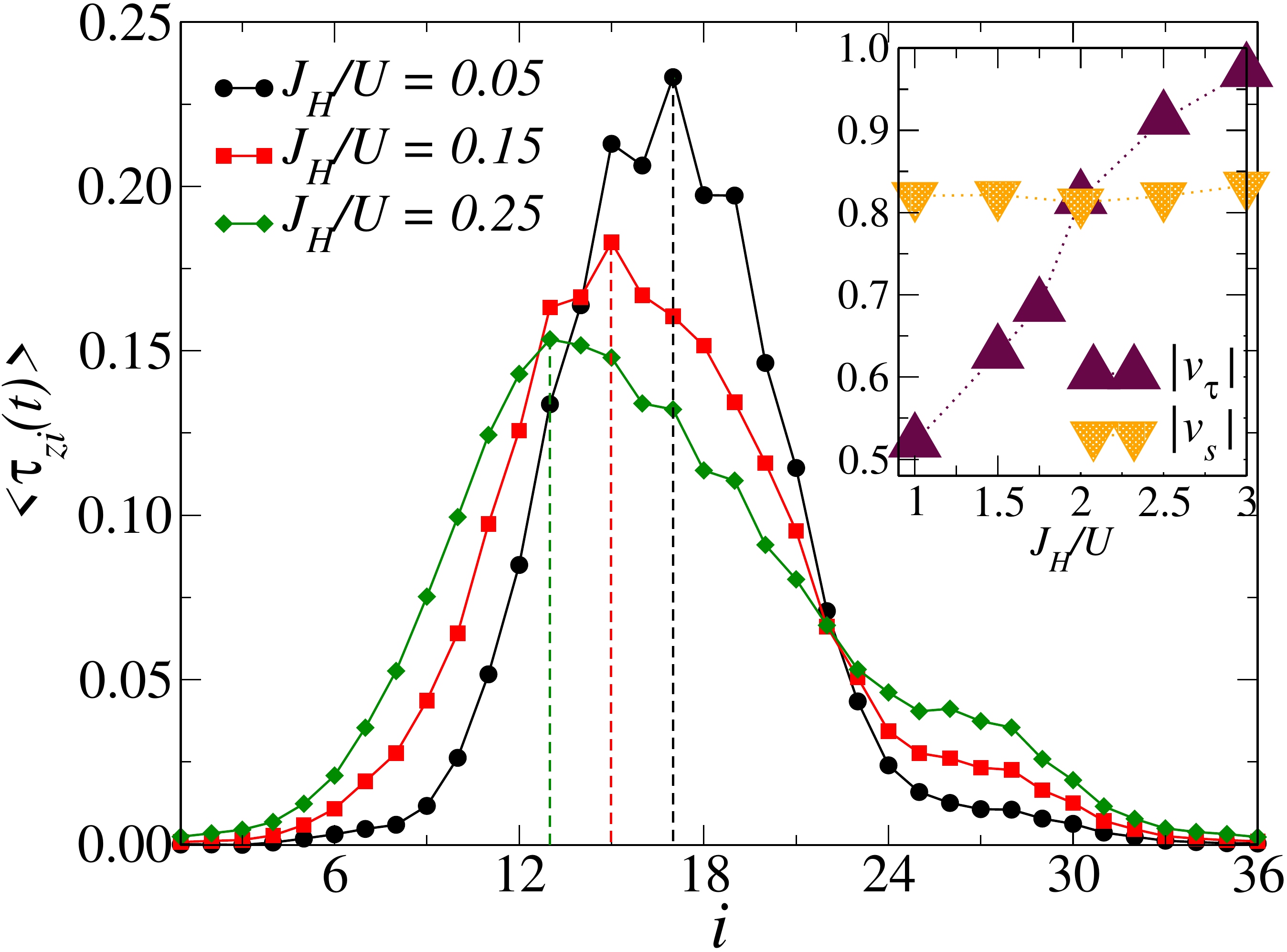}}
	\caption{Orbital density $\langle \tau_{zi}(t) \rangle$ at time $t=5$ 
	for three values of Hund's interactions $J_H/U$ 
	and at fixed $U/W=1.0$. 
	Inset: orbiton and spinon speeds $|v_{\tau}|$ and 
	$|v_{s}|$ parametric with Hund's interaction $J_H/U$ at $U/W=1.0$. 
	}
\label{Fig3}
\end{figure}

\begin{figure}[h]
\centering
\rotatebox{0}{\includegraphics*[width=\linewidth]{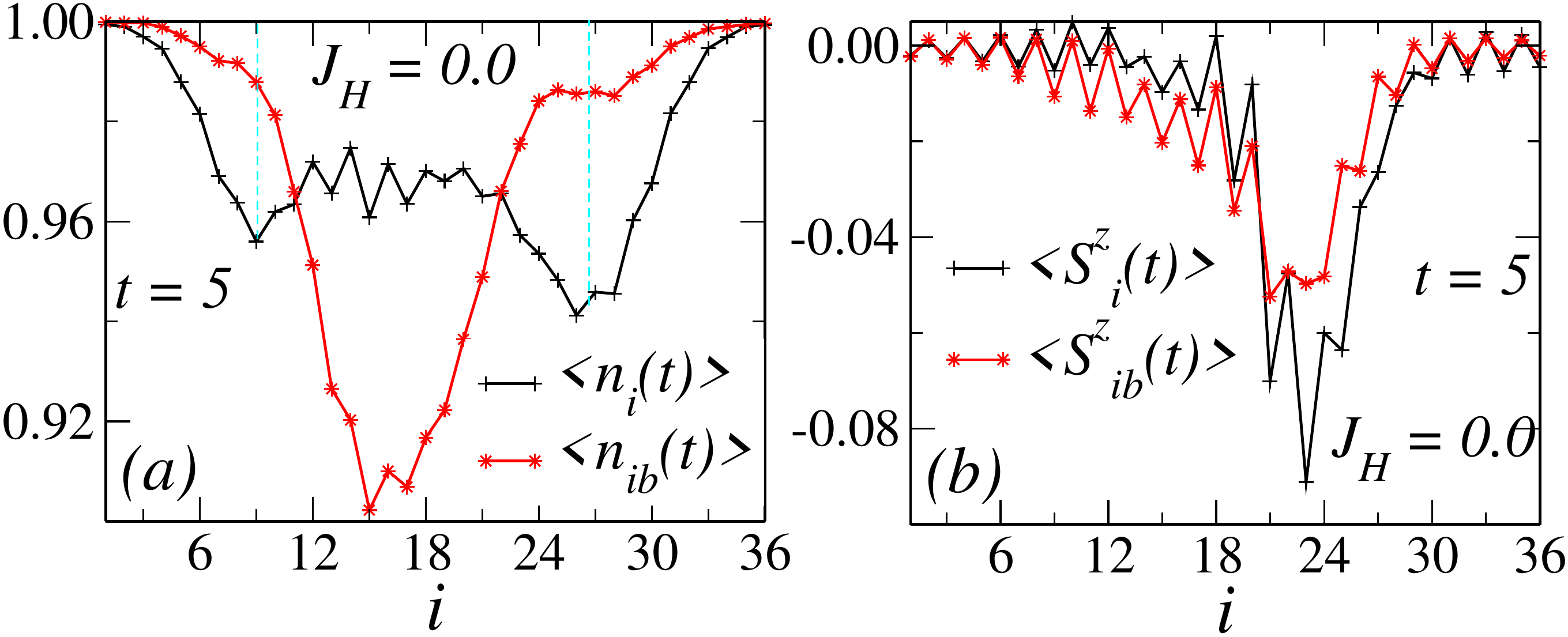}}
	\caption{Comparison of dynamics of a hole in the one-orbital (half-filled) 
	and two-orbital (quarter-filled) 1D chain at $t=5$. 
	(a) Charge densities $\langle n_i(t)\rangle$ (one-orbital)
	and $\langle n_{ib}(t) \rangle$ (two-orbital) 
	(b) Spin densities $\langle S^z_i(t) \rangle$ (one-orbital)
	and $\langle S^z_{ib}(t) \rangle$ (two-orbital).
	}
\label{Fig4}
\end{figure}

Next, to study the role of the inter-orbital repulsion $U'$ 
and Hund's coupling $J_H$ in the dynamics of the exciton wave-packet, 
we calculate $\langle \tau_{zi}(t) \rangle$ and $\langle S^z_{ib}(t) \rangle$
for different values of $J_H/U$. Figure~\ref{Fig3}(a) displays 
the orbital wave-packet at time $t=5$ but for three different values of 
$J_H/U$. We find that for the smaller coupling $J_H/U=0.05$ the
wave-packet $\langle \tau_{zi}(t) \rangle$ traveled only a very short
distance from the central site $i_0=18$. However, increasing 
the Hund coupling to $J_H/U=0.25$, still at time $t=5$, $\langle \tau_{zi}(t) \rangle$ 
traveled a larger distance (five lattice spacings) from the central site $i_0=18$.
 The inset shows the orbiton and spinon speeds $|v_{\tau}|$ and $|v_{s}|$, respectively,
 vs.~$J_H/U$. We find that the orbital velocity increases significantly 
 with increasing $J_H/U$. We believe this is because increasing $J_H/U$ reduces
 the inter-orbital interaction $U'$, which results in a
 less-tightly bound electron-hole pair and, thus, the exciton becomes 
 less heavier and can move at a faster rate $|v_{\tau}|$. On the other
 hand, at small $J_H/U$ the inter-orbital interaction $U'$ 
 increases and results in heavier excitons, which naturally are more
 localized~\cite{yang,shukla}. 
 The larger value of orbiton velocity was observed in RIXS experiments
  because of the large Hund's coupling 
  in Ca$_2$CuO$_3$~\cite{betto}. The increase in orbiton velocity
  was explained in terms of the superexchange process~\cite{hever,betto},
  where they showed that the  energy of the intermediate state
  during the movement of the orbiton depends on Hund's coupling $J_H$. 
  For completeness, note that we find the spin speed $|v_s|$ (inset)
 does not change much with increasing $J_H/U$, and 
 remains unaffected by the concomitant modifications in $U'$, which is intuitively reasonable.


\begin{figure}[h]
\centering
\rotatebox{0}{\includegraphics*[width=\linewidth]{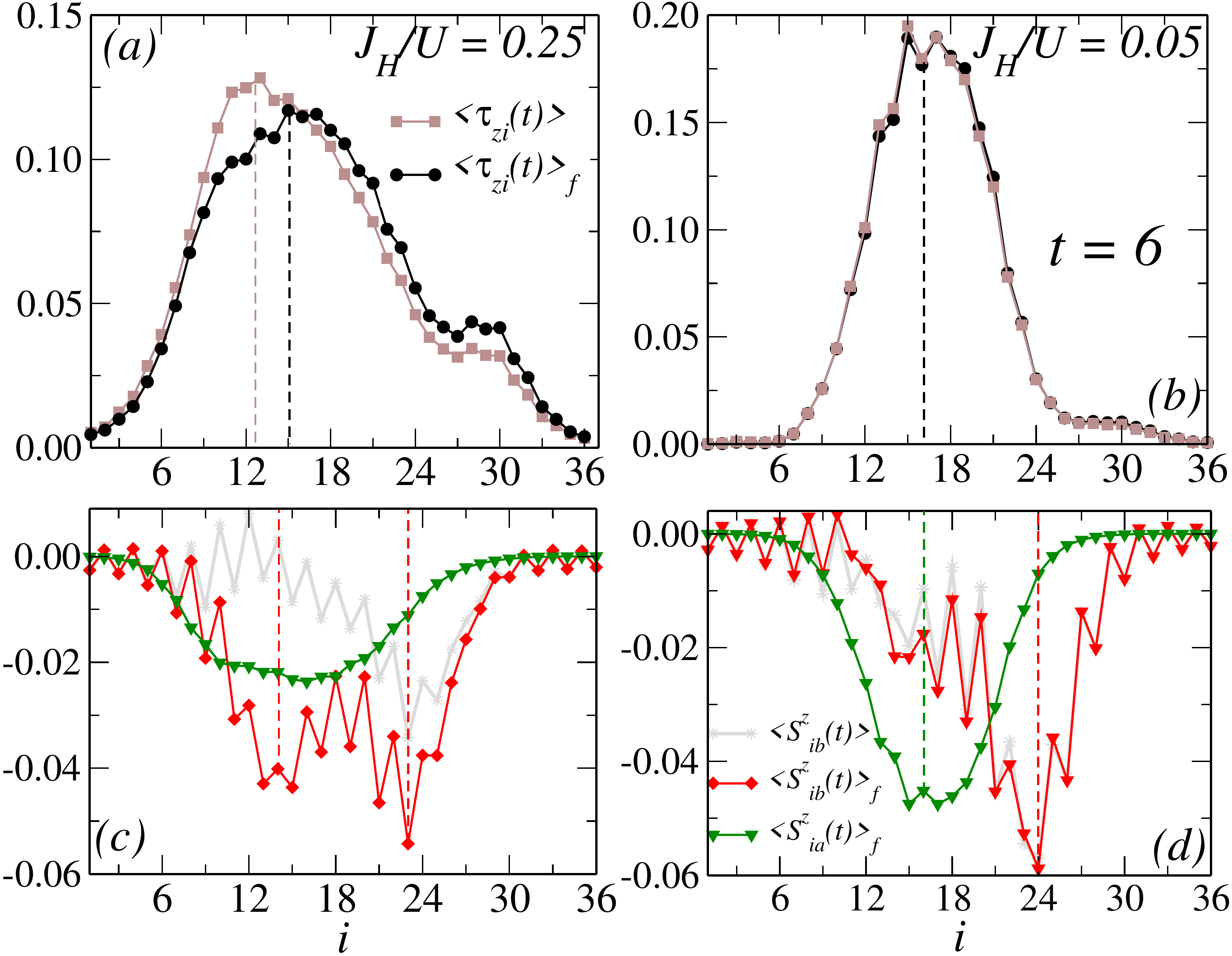}}
	\caption{Comparison of dynamics of exciton with and without
	spin-flip process. The orbital wave-packet with spin-flip (without spin flip)
	 $\langle \tau_{zi}(t) \rangle_f$ ($\langle \tau_{zi}(t)\rangle$) is denoted by circles (squares).
	 (a) is for $J_H/U=0.25$ and (b) for $J_H/U=0.05$. 
	 Spin wave-packets $\langle S^z_{ib}(t) \rangle_f$ (diamonds),
	 $\langle S^z_{ia}(t) \rangle_f$ (down-triangle) with spin-flip process
	and $\langle S^z_{ib}(t) \rangle$ (stars) without spin-flip for 
	(c) $J_H/U=0.25$ and (d) $J_H/U=0.05$. These results were obtained
	at time $t=6$ and for $U/W=1.0$
	}
\label{Fig5}
\end{figure}

In Figs.~\ref{Fig4}(a) and (b), we show a comparison of the dynamics of
a hole in the one-orbital Hubbard model (half-filled chain, $U/W=1.0$) 
and in the two-orbital Hubbard model (quarter-filled, $U/W=1.0$, $U'/W=1.0$, and $J_H/U=0$)
chain system. At $t=0$, a hole was created at the central site $i_0=18$,
either by removing an electron at site $i_0=18$ for the one-orbital case or, for two orbitals, removing
an electron in orbital $b$  and exciting this electron to orbital $a$ at the same site $i_0=18$.
The results for the charge wave-packets are remarkably different. 
While the charge wave-packet in the one-orbital system moves quite fast
and splits into left and right moving  wave-packets,
the charge wave-packet in the two-orbital system moves very slowly
due to the formation of the strong electron (in orbital $a$)-hole (in orbital $b$) bound state. The heaviness of the bound state electron-hole is 
natural because to propagate to the next site, 
it involves two hoppings {\it $t_a$} and {\it $t_b$} and an intermediate state 
with energy proportional to {\it U} [scale as {\it $t_a t_b/U$}], 
while the bare hole in one orbital propagates easily with just a hopping {\it $t_{hop}$}.
Interestingly,
the spin wave-packets in both systems move with a similar speed 
and towards the right from the central site $i_0=18$
[see Fig.~\ref{Fig4}(b)]. This is expected because after the separation 
of spin and charge wave packets in the two-orbital system 
(at quarter-filling), the spin moves approximately guided by the scale $t_{hop}^2/U$, the same as 
the spinon follows in the one-orbital half-filled system~\cite{jagla}. 
\begin{figure}[h]
\centering
\rotatebox{0}{\includegraphics*[width=\linewidth]{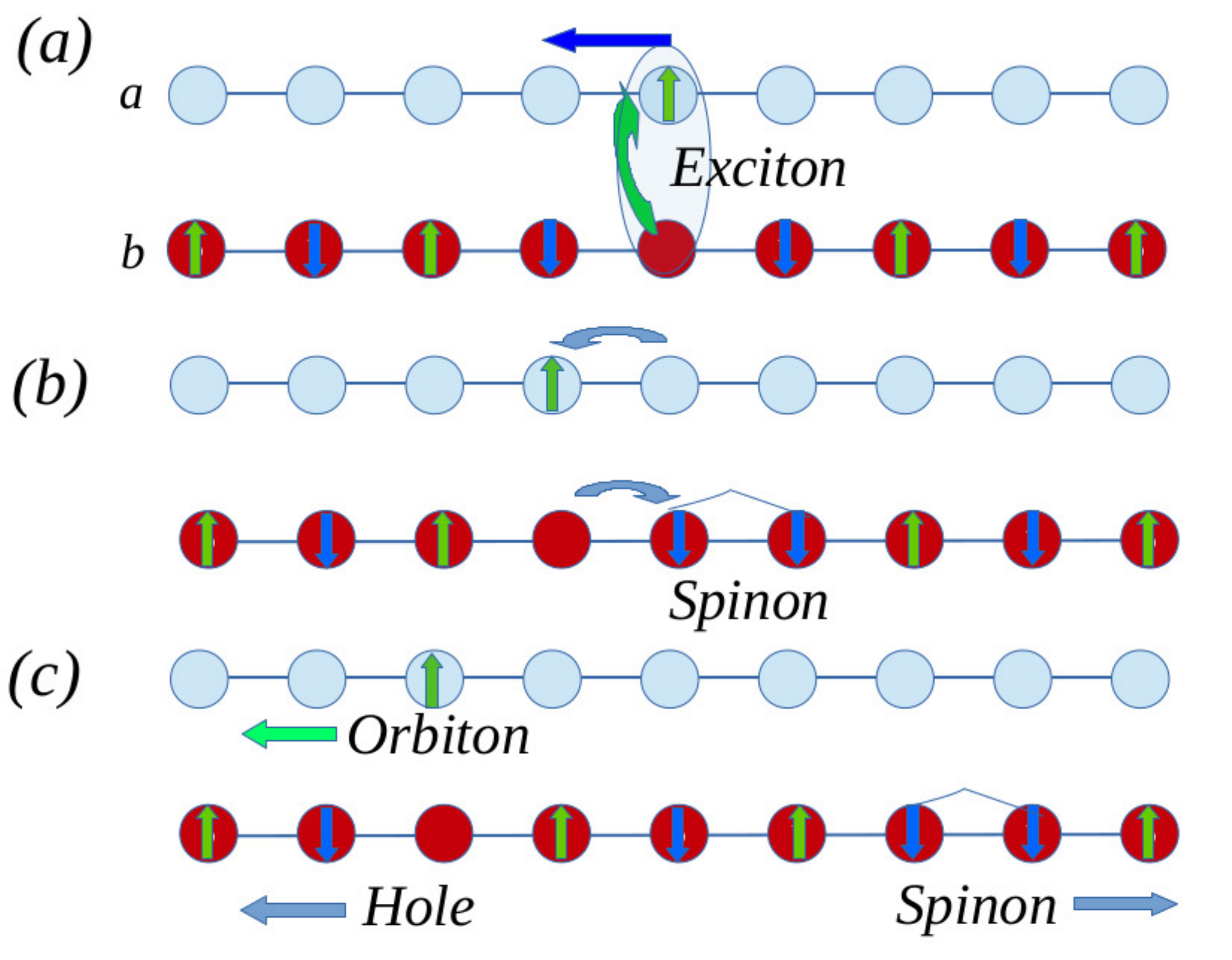}}
	\caption{Illustration of spin-orbit and spin-charge
	separation in a two-orbital ($a$ and $b$) one-dimensional chain.
	(a) An exciton with a finite momentum (blue arrow) is created at $t=0$ by exciting an electron from orbital $b$. 
	(b) Electron in orbital $a$ hops towards the left, while an electron 
	with down spin on orbital $b$ hops towards the right (i.e. hole moves to the left), creating a spinon 
	on orbital $b$. (c) Orbiton and hole move in a bound state to the left, while spinon
	moves free to the right.
	}
\label{Fig6}
\end{figure}

In the RIXS experiment during the creation of orbital excitations
spin-flip processes are also allowed~\cite{hever,betto}.
Figure~\ref{Fig5} presents a comparison of orbiton dynamics 
with and without spin-flip during the exciton generation, for different values of $J_H/U$.
At $t=0$, for the spin-flip process the exciton wave-packet 
was created by the operator 
$A \sum_j e^{-\left(j-j_0\right)^2/2{\omega_r}^2}e^{-ik_0j} c^{\dagger}_{j\downarrow a}c^{\phantom\dagger}_{j \uparrow b}$ 
acting on the ground state wave-function $|\Psi_0 \rangle$.
We found that the orbital velocity of $\langle \tau_{zi}(t) \rangle_f$ when spin flip occurs is only
slightly reduced compared to the previously described 
spin non-flip case $\langle \tau_{zi}(t) \rangle$ at $J_H/U=0.25$ 
[see Fig.~\ref{Fig5}(a)]. The slower speed of the spin-flip orbiton,
compared to the spin-non-flip orbiton, is not expected from the 
superexchange picture~\cite{betto} (where the spin-flip orbiton moves regulated by
$t_at_b/(U-3J_H)$ and for spin non-flip case by $t_at_b/(U-2J_H)$)~\cite{note1}.
For a smaller $J_H/U=0.05$, the results are almost identical and the orbital wave-packet 
moves very slowly in both cases [see Fig.~\ref{Fig5}(b)].

The spin-flip excitonic process leads to the creation of spin wave-packets 
$\langle S^z_{ib}(t) \rangle_f$ and $\langle S^z_{ia}(t) \rangle_f$ 
in the spin-down state [Fig.~\ref{Fig5}(c)].
Interestingly, at large $J_H/U$ the spin-wave packet
splits into two wave-packets with time ($t\gtrsim 3$), 
travelling in opposite directions (starting at the central site $i_0=18$). 
This curious splitting of the spin wave-packet $\langle S^z_{ib}(t) \rangle_f$
indicates the presence of two fractionalized spinons~\cite{chen}
The left moving wave-packet $\langle S^z_{ib}(t) \rangle_f$ travels with similar 
speed as $\langle S^z_{ia}(t) \rangle_f$ of orbital $a$ and 
$\langle \tau_{zi}(t) \rangle_f$. 
This could be due to the strong Hund's interaction between spin wave-packets of 
 orbital $a$ and $b$, which favors parallel alignment (spin-down state) of 
spin wave-packets $\langle S^z_{ib}(t) \rangle_f$ and $\langle S^z_{ia}(t) \rangle_f$. 
At large $J_H/U$, the
 creation of additional spinons was suggested in the spin-orbital spectrum~\cite{hever}. 
In the case of spin-flip excitation, 
 a strong $J_H/U$ also leads to attraction between 
 orbiton and spinon~\cite{hever}, which may be related to  
 the slight slowdown of the orbiton velocity [Fig.~\ref{Fig5}(a)] 
 compared to the without-spin-flip case (where spinon and orbiton repel
 each other~\cite{hever}).
The right moving spin-wave packet  $\langle S^z_{ib}(t) \rangle_f$ moves 
with speed similar to that of $\langle S^z_{ib}(t) \rangle$ (without spin-flip case) [Fig.~\ref{Fig5}(c)]. 
On the other hand, for smaller $J_H/U$, 
the spin wave-packet  $\langle S^z_{ib}(t) \rangle_f$ 
does not split into two parts. $\langle S^z_{ib}(t) \rangle_f$ (spin-flip case)
and  $\langle S^z_{ib}(t) \rangle$ (without spin-flip) move with similar speeds (see Fig.~\ref{Fig5}(d)).

{\it Conclusions.} 
Using the Krylov-space t-DMRG method we 
studied the real-time dynamics of an excitonic wave-packet evolving via a
two-orbital Hubbard model on a chain, at intermediate coupling $U/W$.  
We observed the real-space spin-orbit and spin-charge
separation by monitoring the dynamics of spin, charge, and orbital 
wave-packets. We find that the
charge and spin wave-packets of the higher energy orbital $a$ 
move together, whereas the charge and spin wave-packets of the lower energy
orbital $b$ moves in opposite direction (Fig.~\ref{Fig6}). 
The electron in the higher energy orbital and hole 
in the lower energy orbital always moves together.
The inter-orbital interactions ($U'$ and $J_H$)
play a crucial role in orbiton dynamics. 
For example, the orbiton velocity increases significantly
by increasing $J_H/U$, whereas the spinon velocity remains unchanged. 
Interestingly, we found that \emph{a hole in a one-orbital chain moves
much faster than a hole in a two-orbital chain},
because the hole in the lower energy orbital forms a (heavy) bound pair with
the electron in the higher energy orbital.
Moreover, we presented the dynamics of the spin-flipped exciton, 
where we found evidence of fractional spinons at 
large Hund's coupling. Our calculations will be extended in future 
work in various directions. For example, into other chain multiorbital
systems with exotic states~\cite{con1},
including spin-orbit coupling~\cite{con2}, 
into ladder geometries~\cite{con3}, 
in materials with orbital order~\cite{con4}, ruthenates~\cite{con5}, 
and into generic $t-J$ models~\cite{con6}.

{\it Acknowledgments.} We thank N. Kaushal and N. D. Patel for discussions.
B.P. and E.D. were supported by the U.S. Department of
Energy (DOE), Office of Science, Basic Energy Sciences
(BES), Materials Sciences and Engineering Division.
G.A. was partially supported by the Center for Nanophase Materials Sciences,
which is a U.S. DOE Office of Science User Facility, and by the Scientific Discovery  through  Advanced  Computing  (SciDAC) program  funded  by  U.S.  DOE,  Office  of  Science, Advanced  Scientific  Computing  Research  and  Basic Energy Sciences, Division of Materials Sciences and Engineering.
Validation and some computer runs were conducted at the Center for Nanophase Materials Sciences, which is a DOE Office of Science User Facility.


\begin{thebibliography}{99}
\bibitem{note1}These differences highlight the importance of studying a Hubbard model because the intermediate and strong coupling regimes do not necessarily behave equally.
\end{thebibliography}
\end{document}